# Parametrically Amplified Low-Power MEMS Capacitive Humidity Sensor

R. Likhite, *Member, IEEE*, A. Banerjee, A. Majumder, H. Kim, *Member*, IEEE, and, *C.H. Mastrangelo, Sr. *Member, IEEE*

***Abstract*— We present the design, fabrication, and response of a polymer-based Laterally Amplified Chemo-Mechanical (LACM) humidity sensor based on mechanical leveraging and parametric amplification. The device consists of a sense cantilever asymmetrically patterned with a polymer and flanked by two stationary electrodes on the sides. When exposed to a humidity change, the polymer swells after absorbing the analyte and causes the central cantilever to bend laterally towards one side, causing a change in the measured capacitance. The device features an intrinsic gain due to parametric amplification resulting in an enhanced signal-to-noise ratio (SNR). 11-fold magnification in sensor response was observed via voltage biasing of the side electrodes without the use of conventional electronic amplifiers. The sensor showed a repeatable and recoverable capacitance change of 11% when exposed to a change in relative humidity from 25-85%. The dynamic characterization of the device also revealed a response time ~1s and demonstrated a competitive response with respect to a commercially available reference chip.**

## I. Introduction

In recent decades, miniaturized humidity sensors have been realized using various transduction methods [1]–[7] for a wide range of applications such as improving indoor comfort in homes and automobiles, humidity monitoring in semiconductor processing facilities [8], food processing industries [9], medical facilities [10] and, Internet-of-Things (IoT) based frameworks [11]. The push for the need for low-power chemical sensors [12]–[14] has been quite strong due to the growing importance of IoT sensor nodes across the world.

The most commonly used humidity sensor for the applications mentioned above is the capacitive sensor, which is used in nearly 75% of the cases [3] as it consumes near-zero DC power. These devices measure the change in capacitance caused by variations in dielectric properties or thickness of a sensing layer sandwiched between two parallel plates [15], [16] when exposed to humidity. Sensors based on measuring the deflection of a microcantilever coated with a sensing polymer have also been demonstrated [17], [18]. These sensors show a linear behavior, are easy to batch fabricate and most importantly, consume nearly zero DC power. However, their Signal-to-Noise Ratio (SNR) is usually limited by electronic noise as their sense capacitance can be small in comparison to the surrounding parasitics due to their small size [19]. Amplifiers are typically used in conjunction with such sensors to obtain a stable output which further adds to the total noise of the system and limits the SNR.

The sensitivity of capacitive hygrometers can be significantly improved if these devices have an intrinsic gain, thus reducing the dependence on noisy electronic amplifiers. This can be achieved with mechanical leveraging and parametric amplification. An example of a mechanically leveraged structure is a microcantilever device coated on one side with a sensing layer. Highly sensitive microcantilever-based sensors have been reported previously for detecting gases [18], [20], [21], DNA hybridization [22], [23], and toxic chemical warfare agents [24]. In these devices, the exposure of the sensing film to an analyte generates surface stress that induces bending of the free-standing cantilever either due to a reduction in interfacial surface energy or swelling of the sensing layer. High signal to noise ratio can be realized in such devices using parametric amplification while maintaining small transducer size and low power consumption by exploiting the voltage induced lateral instability in MEMS devices to magnify their displacement to capacitance sensitivity. This technique has been previously reported to improve the performance of a MEMS magnetometer [19], gyroscope [25], hair-flow sensor [26], and vapor sensors [27]–[29]. Unlike electronic amplification, parametric amplification has an inherent advantage of providing higher sensitivity in MEMS sensor systems as it amplifies the sensor signal, without adding any extra electronic noise to the circuit.

In this paper, we report the design, fabrication, and testing of a new type of low power, batch-fabricatable parametrically amplified microcantilever-based humidity sensor with improved sensitivity. This article expands on a proof of concept presented earlier [29]. Extensive characterization of the sensor response has been presented in this article, along with dynamic response testing and comparison to a commercially available sensor. Additionally, a relevant analysis of the sensor output and a mathematical model describing the sensor action is also The authors are with the department of Electrical and Computer Engineering at the University of Utah, Salt lake city, UT, USA.

H. Kim and C.H. Mastrangelo are also with the department of Biomedical Engineering at the University of Utah.

*Corresponding author: C.H. Mastrangelo (email: carlos.mastrangelo@utah.edu)

presented along with a study on the sorption kinetics of the device.

## II. Device structure and operation

A unit cell of the LACM sensor is shown in Fig. 1a. The device consists of a suspended microcantilever beam (electrode 2) asymmetrically coated on top with a sensing polymer (polyimide) and flanked on the sides by two stationary electrodes (1 and 3). When the device is exposed to an analyte vapor, the polyimide absorbs the gas and swells. This exerts a bending moment, $M_{\%RH}$, on the structural beam causing it to deflect to one side (Fig. 1b). Unlike conventional polymer-based microcantilever sensors which measure the out-of-plane deflection of the cantilever [30], the LACM sensor measures the in-plane deflection of the sensing cantilever by forming two parallel plate variable capacitors between the central finger and the adjacent electrodes as shown below. Our devices are appropriately designed to make the out-of-plane stiffness much higher than the in-plane spring constant. A planer design allows multiple unit cell structures to be ganged up in parallel to increase the total output signal from the sensor (Fig. 1c) while maintaining compatibility with conventional CMOS processes suitable for low power, high sensitivity water vapor sensor for application in IoT frameworks.

Furthermore, parametric amplification of the output signal is achieved by applying a symmetric DC bias voltage to both the flanking electrodes (1 and 3) w.r.t the central suspended electrode (2) to improve the vapor-concentration to displacement sensitivity of the device (Fig. 1d). In this work, the single side capacitance measurements for the device have been reported. Theoretically, the sensor performance can be further improved by measuring a differential capacitance between the two sides while also eliminating common mode parasitics.

### A. Electrostatic spring softening

Parametric amplification induced spring softening in the mechanical domain has been extensively studied to tune the resonant frequency of MEMS structures [31]–[33] and produce large-amplitude deflections in microstructures [26], [34]. In the LACM sensor, when a DC bias voltage is applied to the electrodes 1 & 3 w.r.t electrode 2, the non-linearity of the electrostatic forces acting on the central cantilever beams results in the reduction of the effective spring constant of the central cantilever. Mathematically, electrostatic spring softening of micromechanical systems can be observed by minimizing the total energy ($U_T$) function of the system,

$$U_T = U_{EL} + U_M \tag{1}$$

Where $U_{EL}$ is the electrostatic energy stored in the capacitors of the system and $U_M$ is the mechanical energy stored in the deformed microcantilever beam. For a microcantilever beam deflecting laterally between two electrodes, as shown in Fig. 1, the respective energies can be written as,

$$U_{EL} = \frac{-\varepsilon A V_b^2}{2}\left(\frac{1}{g_0+\Delta y} + \frac{1}{g_0-\Delta y}\right) \tag{2}$$

$$U_M = \frac{k_o \Delta y^2}{2} \tag{3}$$

Where, $g_0$ is the initial gap between the electrodes and $\Delta y$ is the deflection of the central beam due to absorption induced polymer swelling, $V_b$ is the applied DC bias voltage, $A$ is the overlap area of the capacitor, $\varepsilon$ is the permittivity and $k_o$ is the lateral spring constant of the central finger when no bias is applied. For $\frac{\Delta y}{g_0} \ll 1$, the total energy of the system can be written as,

$$U_T = \frac{1}{2}k_o\Delta y^2 - \left[\frac{\varepsilon A V_b^2}{g_0}\right]\left(\frac{1}{1-\left(\frac{\Delta y}{g_0}\right)^2}\right) \tag{4}$$

$$U_T = \frac{1}{2}k_o\left[\Delta y^2 - \left(\frac{2\varepsilon A V_b^2}{k_o g_0}\right)\left(1+\left(\frac{\Delta y}{g_0}\right)^2 - \left(\frac{\Delta y}{g_0}\right)^4 + \dots\right)\right] \tag{5}$$

$$U_T \cong \underbrace{\frac{1}{2}k_o\left[1-\left(\frac{2\varepsilon A V_b^2}{k_o g_0^3}\right)\right]}_{k(M)}\Delta y^2 + h.o.t \tag{6}$$

For $V_p = \sqrt{\frac{k_o g_0^3}{2\varepsilon A}}$, which is the differential pull-in voltage for the structure, the effective softened spring constant $k(M)$, can be written as,

$$k(M) = \frac{k_o}{M}\ where, M = \frac{1}{\left(1-\frac{V_b^2}{V_p^2}\right)} \tag{7}$$

For example, Fig. 2a shows the total energy of a system with $k_y = 3.7\ Nm^{-1}$ and $V_p = 35.4V$ without DC bias ($V_1$ = 0V) and, when DC voltage bias ($V_2$ = 18V and $V_3$ = 28V) is applied to induce parametric amplification, as a function of normalized beam deflection. As the magnitude of the applied voltage bias is increased, the system becomes progressively unstable due to shallowing of the local energy minima. Since $k(M) < k_o$, the output signal is magnified by a voltage-dependent magnification factor, M, as shown in Fig. 2b.

### B. Noise Analysis and Signal-to-Noise Ratio

The noise of the system originates from both the mechanical-thermal noise [35], [36] of the deflecting beam and the noise of the C/V converter operational amplifiers [37]. In our device, we convert the stress caused by the RH absorption into a mechanical deflection which translates into a variable, RH dependent capacitance, mediated through the beam spring constant $k(M)=k_o/M$, where $M$ is the bias-dependent magnification factor. We use high-gain op-amps amplify and read the sensor capacitance [38] as illustrated in the schematic of Fig. 3a below.

The capacitance change indicative of the vapor concentration signal is first converted to an AC current $i_{ac} = j\omega C(RH)\cdot V_o$ , where $V_o$ is the amplitude of a sinusoidal AC voltage of angular frequency $\omega$ placed across the capacitor. The AC current is finally converted to an output voltage using a high-gain op-amp of transconductance $G_o$. The noise of the system is calculated by the introduction of two noise sources: (1) a mechanical noise acting on the beam with spectral density [35], [39],

$$F_B = \sqrt{4k_BTR} \qquad [N/\sqrt{Hz}] \tag{8}$$

Which, is only dependent on the Boltzmann constant $k_B$, temperature $T$, and, viscous damping coefficient $R$, of the system, as described previously in a viscous damping environment (air) and, at low pressures [36]. Therefore, the noise force on the MEMS device is completely independent of the bias induced spring-softening used in the LACM sensor. The second electronic noise current source $\overline{i_e^2}$ is introduced at the input of the transconductance amplifier. The output noise for this system is thus

$$\overline{v_o^2\ (k_o)} = G_o^2 \left( \omega^2 V_o^2\, F^2 \frac{\overline{F_B^2}}{k(M)^2} + \overline{i_e^2} \right)$$

$$= G_o^2 \left( \omega^2 V_o^2\, F^2 \frac{\overline{F_B^2}}{k_o^2} + \overline{i_e^2} \right) = G_o^2 \left( H^2(\omega) \cdot \frac{\overline{F_B^2}}{k_o^2} + \overline{i_e^2} \right) \quad (9)$$

Where

$$H^2(\omega) = \omega^2 V_o^2\, F^2 \quad , \; F^2 = \left(\frac{dC}{dF_B}\right)^2 \quad (10)$$

And since the deterministic signal $v_{RH}^2 = H^2(\omega) \cdot \frac{F_{RH}^2}{k_o^2} \cdot G_o^2$, where $F_{RH}$ is the equivalent RH-driven force, the signal to noise ratio is thus

$$SNR_o = \frac{\frac{F_{RH}}{k_o} \cdot H(\omega)}{\sqrt{\frac{F_B^2}{k_o^2} \cdot H^2(\omega) + i_e^2}} = \frac{\left(\frac{F_{RH}}{F_B}\right)}{\sqrt{1 + \left(\frac{k_o^2}{H^2(\omega) \cdot F_B^2}\right) \cdot i_e^2}}$$

$$= \frac{SNR_{max}}{\sqrt{1 + \left(\frac{k_o^2}{H^2(\omega) \cdot F_B^2}\right) \cdot i_e^2}} \quad (11)$$

It is evident that the $SNR$ is independent of transconductance $G_o$. Eq. (11) Also tells us that any electromechanical effect that lowers $k_o$ will result in a higher $SNR$. In our device, this can be achieved with the bias-induced spring softening gain $M$, as shown in the schematic of Fig. 3b such that

$$SNR(M) = \frac{SNR_{max}}{\sqrt{1 + \frac{\left(\left(\frac{SNR_{max}}{SNR_o}\right)^2 - 1\right)}{M^2}}} \quad (12)$$

The SNR is improved for $M > 1$ essentially because the spring softening effect gain is noiseless, thus moving the electronic noise closer to the amplified output.

### C. Sensor response model

The LACM sensor is essentially a humidity-dependent variable parallel plate capacitor. When the device is exposed to humidity, the sensing polymer swells after absorbing the water vapor. Since the polymer is asymmetrically patterned and constrained to the silicon cantilever beam, the swelling generates surface stress [40], which results in the bending of the cantilever beam towards one side. This results in a change in the measured capacitance of the device. The amount of cantilever bending and, therefore, the capacitance change is directly proportional to the swelled induced surface stress and inversely proportional to the effective spring constant of the device given by equation (7). This can be mathematically described by a modified form of Stoney's equation as given by Godin et al. [41].

For small displacements of the central cantilever beam, the normalized change in capacitance can be written as,

$$\frac{\Delta C}{C} = (A_o) \left[ \frac{(1-\nu) l^2}{g_o E_{Si} w^2} \cdot \frac{1}{\left(1 - \frac{V^2}{V_p^2}\right)} \cdot \beta_p \cdot C_{RH} \right] + B_o \quad (13)$$

Where, $\nu$ = Poisson's ratio of Silicon, $l$ = length of central cantilever beam, $g_0$ = initial air-gap between electrodes, $E_{Si}$ = Young's Modulus of Silicon, $w$ = width of central cantilever beam, $V$ = applied voltage bias to induce spring softening, $V_p$ = differential pull-in voltage of beam, $\beta_p$ = fitting parameter proportional to the swelling-induced surface stress generated by the polyimide and $C_{RH}$ is the relative humidity of the chamber. $A_o$ and $B_o$ are dimensionless fitting parameters.

### D. 2-level electrical interconnects

The laterally deflecting and planar design of the LACM sensor allows connecting multiple unit cells in parallel to further increase the output of the sensor. Since each microcantilever is flanked on the two sides by electrically isolated anchored electrodes, it is necessary to have a 2-level electrical connection arrangement in the device. In the LACM sensor array, this is done by fabricating jumpers out of doped poly-Si, as shown in Fig. 1c. The detailed fabrication procedure is described in the following sections. The jumper arrangement eliminates the needs for wire bonds to connect multiple devices, thus keeping the fabrication process simple.

## III. Fabrication and Imaging

### A. Device Fabrication

Fig. 4(a-l) shows a simplified fabrication procedure of the device. The process starts by depositing 250nm of low-stress silicon nitride using LPCVD process over SOI wafers with 30 µm thick device layer and 2 µm thick buried oxide layer, as shown in Fig. 4a. The nitride layer is patterned using conventional UV photolithography with and then etched using $CF_4/O_2$ RIE. The device layer of the SOI wafer is then etched using DRIE to form the fingers (Fig. 4b). A low-frequency RF source (380 kHz) is used for this process to avoid footing and prevent premature release of the device. The photoresist is then removed using Acetone, and a pre-furnace clean is performed. A 100nm thin layer of LPCVD silicon nitride is then deposited on the wafer, and a blanket $CF_4/O_2$ RIE etch is done. This ensures that the nitride remains only on the sidewalls of the etched fingers (Fig. 4c). A 4µm thick layer of sacrificial LPCVD PSG is then deposited on the wafer and annealed in an $N_2$ environment at 1050°C to reflow the PSG. The thickness of the PSG is then reduced to ~2 µm using a blanket RIE etch on the wafer. This deposition-reflow and etch back process is repeated until the etched gaps between different fingers are entirely sealed (Fig. 4d) due to the cusping effect in an LPCVD process thus allowing further processing of the wafer [42], [43]. The sacrificial PSG and the underlying nitride are then patterned using photolithography and RIE to create anchors (Fig. 4e) for poly-Si jumpers and the anti-stiction micro-staple pins. A 4 µm thick layer of poly-Si is then deposited using

LPCVD process. This layer is then doped using phosphorus solid-source doping and annealed at 1050°C for 2 hours (Fig. 4f). A 200nm thick layer of Cr is then deposited over the wafer using DC-sputtering and patterned using a wet etchant to form the metal contact and jumpers (Fig. 4g). The wafer is the then cleaned, and the anti-stiction features are patterned (Fig. 4h). This step also forms the poly-Si jumpers and contact pads, with the previously patterned Cr metal acting as an etch mask. The PSG on the central finger is then patterned and etched to create windows (Fig. 4i) to allow deposition and anchoring of the sensing polymer to the device. We use HD-4104 polyimide [44] as a water vapor sensing material. The polyimide was spin-coated on the sample and then cured in at 300°C in an $N_2$ environment for 3 hours. An adhesion promoter, VM-651, was applied to the sample before spin-coating to improve the adhesion of the polyimide to the substrate and prevent any delamination during the BOE release procedure. A 200nm thick Al layer is then sputtered on the polyimide and then patterned using wet etching to act as a hard mask. The polyimide is then etched using $O_2$ plasma (Fig. 4j) in an Oxford 100 ICP etcher. The devices are then diced, and the chips are released in BOE for 160 mins with constant stirring (Fig. 4k). The chips are rinsed thoroughly in DI water followed by methanol and allowed to air dry.

### B. *Stiction Suppression*

Stiction is one of the main modes of failure in MEMS devices [45]. Device failure due to stiction occurs when suspended MEMS structures such as cantilevers, plates, or beams adhere to the substrate or adjacent features due to lack of sufficient restoring force when subjected to strong capillary forces. Typically, capillary forces arise during device fabrication and cleaning due to the surface tension of water when the sample is allowed to dry. Various methods have been previously used to prevent stiction due to surface tension [45]–[47]. In the LACM device, a different method which utilizes ‘micro staple-pins,’ that hold the released cantilever in place during a wet release procedure has been utilized. The staples are made out of thin poly-Si which can be dry etched, thus eliminating the need for complex and expensive anti-stiction processes.

A very short $SF_6$ etch is finally performed on the devices to etch away the anti-stiction micro-staples and release the fingers (Fig. 4l). Any unwanted etching on the side wall of the fingers is prevented by the thin $Si_3N_4$ film that was deposited over the finger before PSG sealing.

### C. *Imaging*

High-resolution Scanning Electron Microscope (SEM) imaging of the device was done on an FEI Quanta 600 SEM at an accelerating voltage of 20.0 kV to verify the fabricated device structure, as shown in Fig. 5(a-c). Fig. 5a shows the fabricated device array with the poly-Si jumpers acting as the 2nd level of electrical connections shown in the zoomed in image (Fig. 5b). Only the central finger is released during a timed BOE wet etch as the flanking electrodes are much wider and therefore, stay anchored. Zoomed-in image of the central finger is visible in Fig. 5c showing the asymmetrically patterned polyimide on the movable beam and the anti-stiction micro-staple pins clamping the central finger to the side electrodes after wet release. The fabricated devices were 900 μm long, and the suspended beams were 13.3 μm wide. The air-gap measured between the suspended beam and the side electrode was 4.75 μm. The differential pull-in voltage was calculated to be ~29.5V. The fabricated device had 15 unit cells connected in a parallel circuit.

## IV. Testing and Characterization

### A. *Test setup*

The sensor electrical testing was done at a probe station enclosed in a metallic box to create a localized environment for vapor testing. The enclosing box was grounded to reduce outside interference and noise during measurement. The device capacitance was measured using a Keithley 4200A-SCS CVU connected to the probe station at 1MHz frequency using a 30mV AC signal. The noise floor for the test setup was measured to be ~363aF, and the base capacitance of the device was 270fF. The chamber was flushed with dry $N_2$ gas before testing, and a commercial humidifier placed outside was used to humidify the chamber. The relative humidity (RH) of the chamber was monitored using a commercially available BME-280 [48] chip connected to an Arduino Uno board which reported the chamber humidity to a computer. Dehumidification of the chamber was done by purging the test chamber with dry $N_2$ while evacuating the chamber using an in-house vacuum line.

### B. *Sensor action and humidity response*

Fig. 6a shows the normalized response of the sensor as a function of varying water-vapor concentration at different biasing voltages. Sensor capacitance decreases when operated at no-bias voltage, which we believe, is due to a reduction in overlap area because of undesirable out-of-plane downward deflection of the central cantilever beam when exposed to increasing humidity. Application of a small DC bias voltage results in stiffening of the out of plane spring constant of the central finger due to induced electrostatic levitation as described by Tang et al. [49] which prevents out-of-plane deflection. Fig. 6b shows the sensor characteristics at different relative humidity levels as a function of varying bias voltage indicating sensor output amplification as the bias voltage is increased at a constant humidity level. A ~11-fold magnification of sensor response was observed for a bias voltage of 28V compared to when a low bias (5V) voltage was used at 40.83 %RH, as shown in Fig. 7. Fig. 8 shows the dynamic response of the device when exposed to a relative humidity change from 20-90% and operated at a bias voltage of 28V. Highly repeatable device performance was observed, and no sensor saturation was seen, as shown in Fig. 8a. Fig. 8b shows the continuous operation of the sensor over five cycles of humidification/dehumidification and shows near zero baseline drift.

### C. Model Fitting

Fig 9a shows the equivalent electrical circuit of the LACM sensor. The humidity response of the device was fitted to equation (13), and the plot is shown in Fig. 9b. Parameter extraction revealed the mean value of $\beta_p$ ~ 0.114 mN.m$^{-1}$ per ppm of water vapor and the value of fitting parameters $A_0$ and $B_0$ ranged from 2.5 to 12 and -0.06 to -0.01, respectively. The root-mean-square error was found to be 0.1%, 0.39% and, 0.52% for applied bias voltages 5V, 20V and 28V, respectively.

### D. Absorption-Desorption Kinetics

The dynamic response of the device is dependent on the moisture absorption induced swelling of the polyimide on the central cantilever beam. This type of behavior can be explained by a modified version of Fick's second law of diffusion. The performance of the LACM sensor can be effectively modeled using the model as described by Sikame Tagne [50]. The desorption kinetics of the sensor is considered as gradual desorption of water molecules back into the atmosphere and modeled using the Polanyi-Wigner equation [51]. The curve fitted normalized change in capacitance is shown in Fig. 10a. Additionally, Fig. 10b compares the normalized device response to that of a commercially available BME-280 [48] reference sensor chip. It can be observed from the plot that the LACM sensor accurately follows the response of the reference chip which has a response time of 1s [48] with both the sensors reaching their maximum output at the same time.

## V. Conclusion

We presented the design, fabrication, and response of a batch-fabricated capacitive polymer-based humidity sensor based on mechanical leveraging and parametric amplification. The device exploits the electrostatic lateral instability of MEMS structures to achieve a noiseless intrinsic gain, which helps in achieving a better SNR. A ~11-fold magnification in sensor output was achieved by applying a 28V DC bias voltage to the device at constant water vapor concentration due to spring softening. We demonstrate an unassisted and completely recoverable change of 11% in capacitance value when subjected to a humidity change from 25-85%. The dynamic response of the sensor was also characterized, and the sensor showed a comparable response to a commercially available reference chip with ~1s response time. A mathematical model to accurately describe the sensor action has also been presented. Such a sensor is an excellent candidate for low power, low cost, and sensitive vapor-sensor for applications in IoT based frameworks.

## Acknowledgment

The authors would like to thank the staff at the University of Utah Nanofab and, the Surface Analysis Lab for their assistance in fabricating and imaging the LACM sensor.

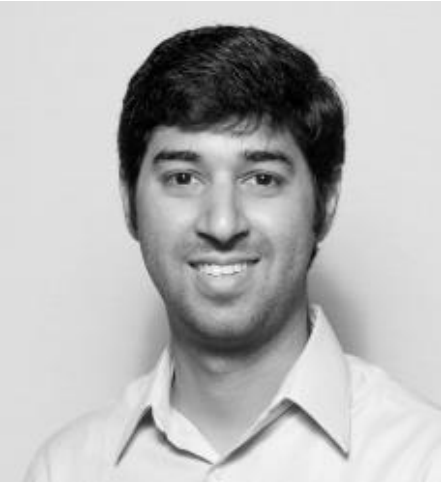

**Rugved Likhite** was born in Nagpur, India in 1992. He received his B.S. degree in Metallurgical Engineering and Materials Science from the Indian Institute of Technology, Bombay in 2014. He was awarded a master’s degree in Electrical and Electronics Engineering in 2018 and, is currently pursuing his Ph.D. at the University of Utah. His research interests include MEMS sensors and actuators, micro and nano-fabrication, device characterization and, process development.

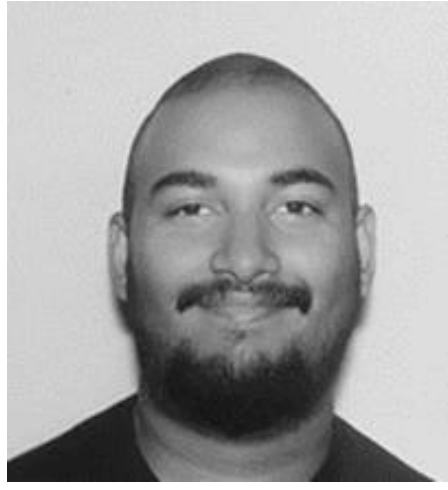

**Aishwaryadev Banerjee** was born in Kolkata, India in 1988. He received his bachelor’s degree in Electronics and Communication Engineering from the West Bengal University of Technology in 2011. In 2013 he joined the Department of Electrical and Computer Engineering at the University of Utah while working as a Research Assistant under Dr. Carlos H. Mastrangelo. Aishwaryadev was awarded a master’s degree in 2018 and his Ph.D. degree in 2019. His research area includes multi-physics simulations, micro and nano-fabrication, MEMS, microsensors, microactuators, and metamaterials, nanoscale physics, nano-scale devices and applied quantum tunneling devices.

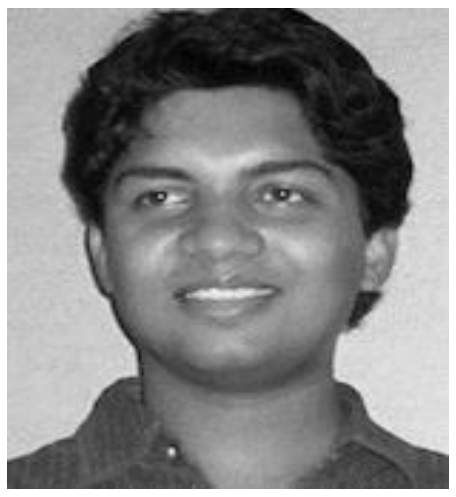

**Apratim Majumder** received his M.S. and Ph.D. degrees in Electrical and Computer Engineering from the University of Utah in 2014 and 2016, respectively. He is currently a Postdoctoral Scholar at the University of Utah. His research interests and areas of expertise include optics and photonics. He is currently responsible for the development of novel nanophotonic device design methodologies, experimental characterization of multi-spectral imaging based on diffractive optical elements, the development and design of multi-level diffractive flat lenses for broadband applications and the research and development of tunable focus liquid-filled lenses for applications in adaptive optics.

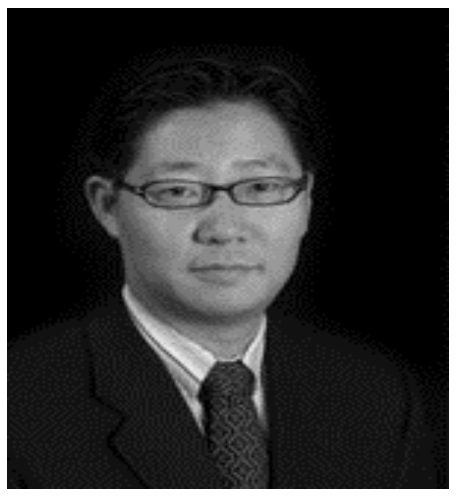

**Hanseup Kim** is an USTAR Associate Professor of Electrical and Computer Engineering, of Mechanical Engineering, and of BioEngineering at the University of Utah in Salt Lake City, Utah since 2009. He is the Director for the Utah Nanofabrication Facility since 2018. He received his BS degree in Electrical Engineering from Seoul National University in 1997, and his MS and Ph.D. degrees in Electrical Engineering from the University of Michigan in 2002 and 2006, respectively. Between 2006 and 2009, he remained as a post-doctoral research fellow at the Center for Wireless Integrated MicroSystems (WIMS) in the University of Michigan working on a micro gas chromatography system, energy harvesting devices, micro hydraulic actuators, and a micro cryogenic cooler. His present research at the University of Utah focuses on the development of integrated intelligent microsystems including chemical and bio sensors, micro actuators, microfluidics, and heterogeneous cell interaction.

Prof. Kim is a recipient of both DARPA Young Faculty Award (YFA) 2011 and NSF CAREER Award 2012. He received some Best Paper Awards or Finalists over the years from the IEEE Conference on Microelectromechanical Systems in 2019, the International Conference on Commercialization of Micro and Nano Systems in 2008 and 2013, and the 38th International Design Automation Conference in 2001. He is also a recipient of the Rotary Club Scholarship in 1999. He has been actively serving the MEMS community as a Technical Program Committee (TPC) member in some conferences and workshops, including the International IEEE Microelectromechanical Systems (MEMS) Conference, the Hilton-Head Solid-State Sensors, Actuators and Microsystems Workshop and PowerMEMS.

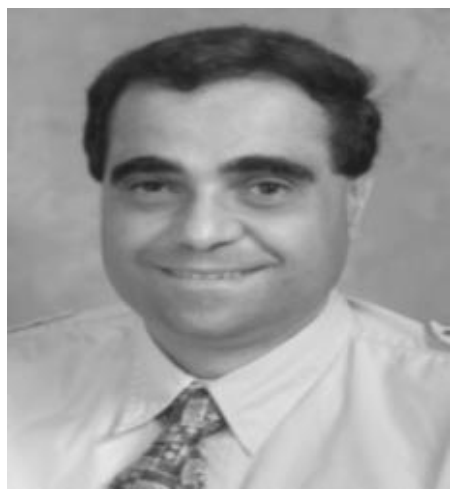

**Carlos H. Mastrangelo** (S’84-M’90) was born in Buenos Aires, Argentina, in 1960. He received the B.S., M.S., and Ph.D. degrees in electrical engineering and computer science from the University of California, Berkeley, in 1985, 1988, and 1991, respectively. His graduate work concentrated on the applications of microbridges in microsensor technology. From 1991 through 1992, he was at the Scientific Research Laboratory, Ford Motor Company, Dearborn MI, developing microsensors for automotive applications. From 1993-2002, he was an Associate Professor of Electrical Engineering and Computer Science at the Center for Integrated Microsystems, University of Michigan, Ann Arbor. From 2000-2005 he was Vice President of Engineering at Corning-Intellisense, Wilmington MA and a Director at the Biochemical Technologies research group, Corning NY. From 2005-2008 he was an Associate Professor of Electrical Engineering and Computer Science at Case Western. He is now a USTAR Professor of Electrical Engineering and Bioengineering at the University of Utah, Salt Lake City. His research focuses on microelectromechanical system applications and technology, microfluidic systems, and integration, design, and modeling of MEMS fabrication processes. Dr. Mastrangelo is on the editorial boards of Sensors and Actuators and the IEEE/ASME *Journal of Microelectromechanical Systems*, and he has participated in technical and organizing committees of numerous SPIE and IEEE conferences in the MEMS area.

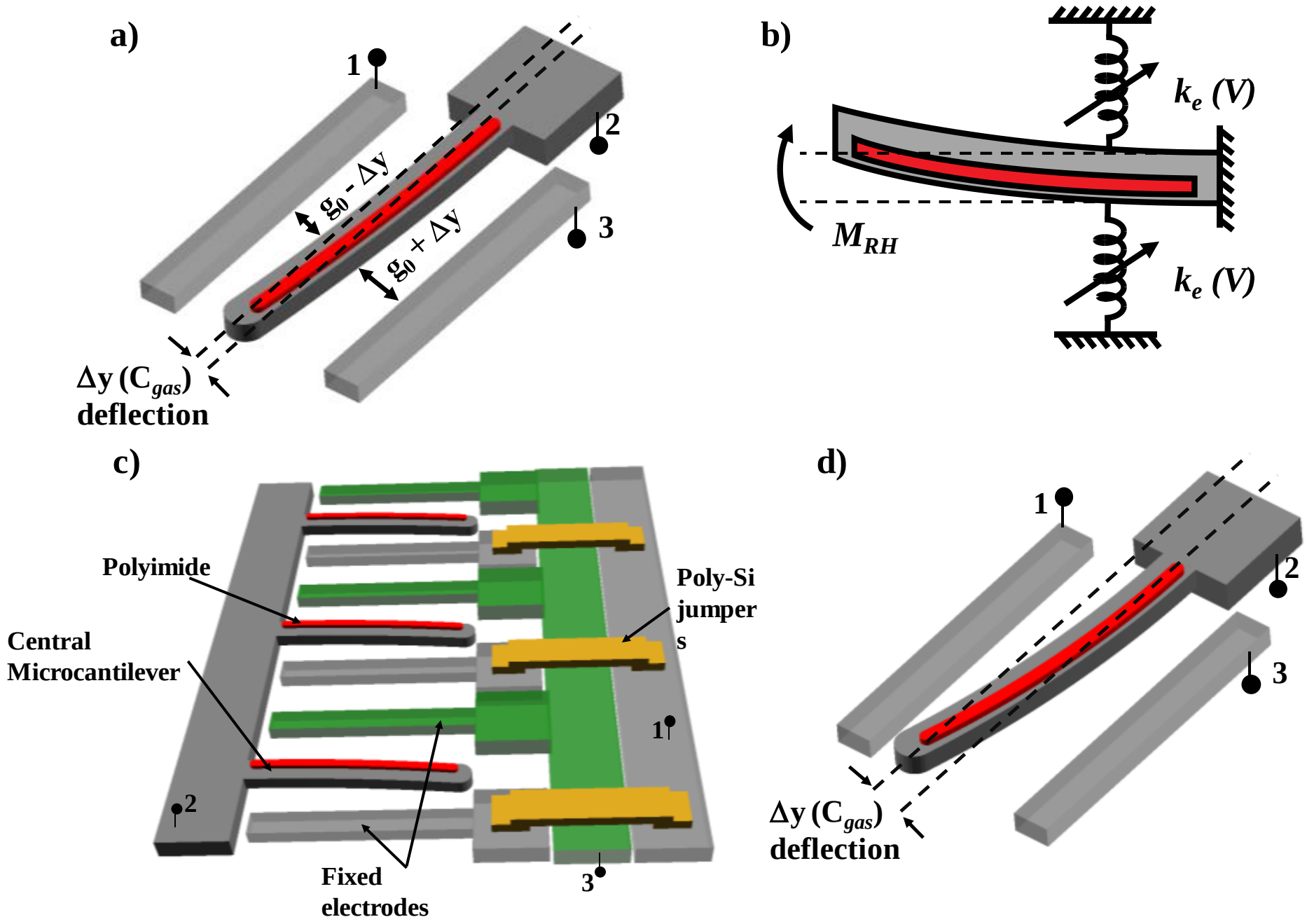

**Figure 1:** a) Unit cell of the LACM humidity sensor. b) Mechanical equivalent of the device when subjected to a change in relative humidity. c) Array structure of parallel LACM unit cells. d) Use of parametric amplification to magnify the deflection sensitivity of the central microcantilever.

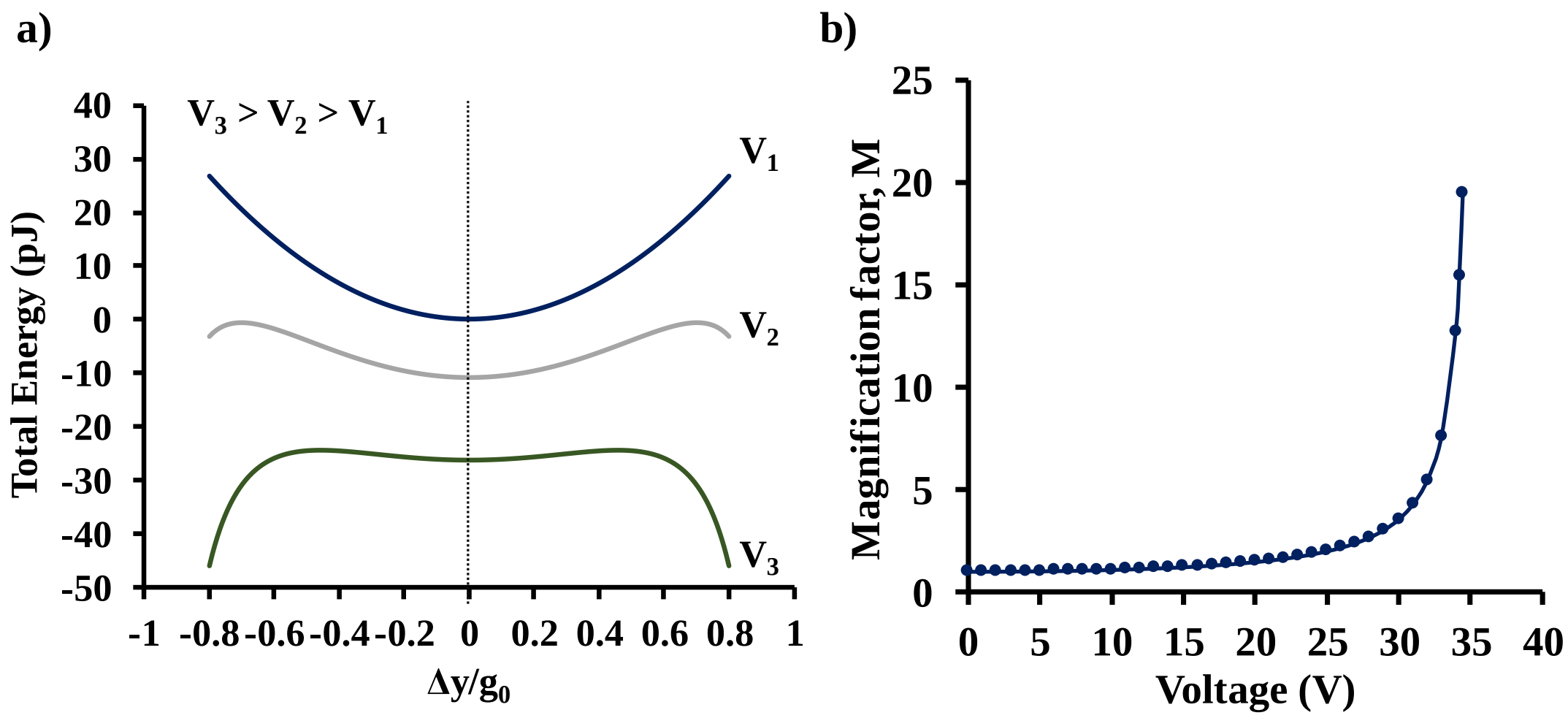

**Figure 2:** Schematic of the a) Total energy of the system during parametric amplification, b) Magnification factor, M as a function of the applied voltage.

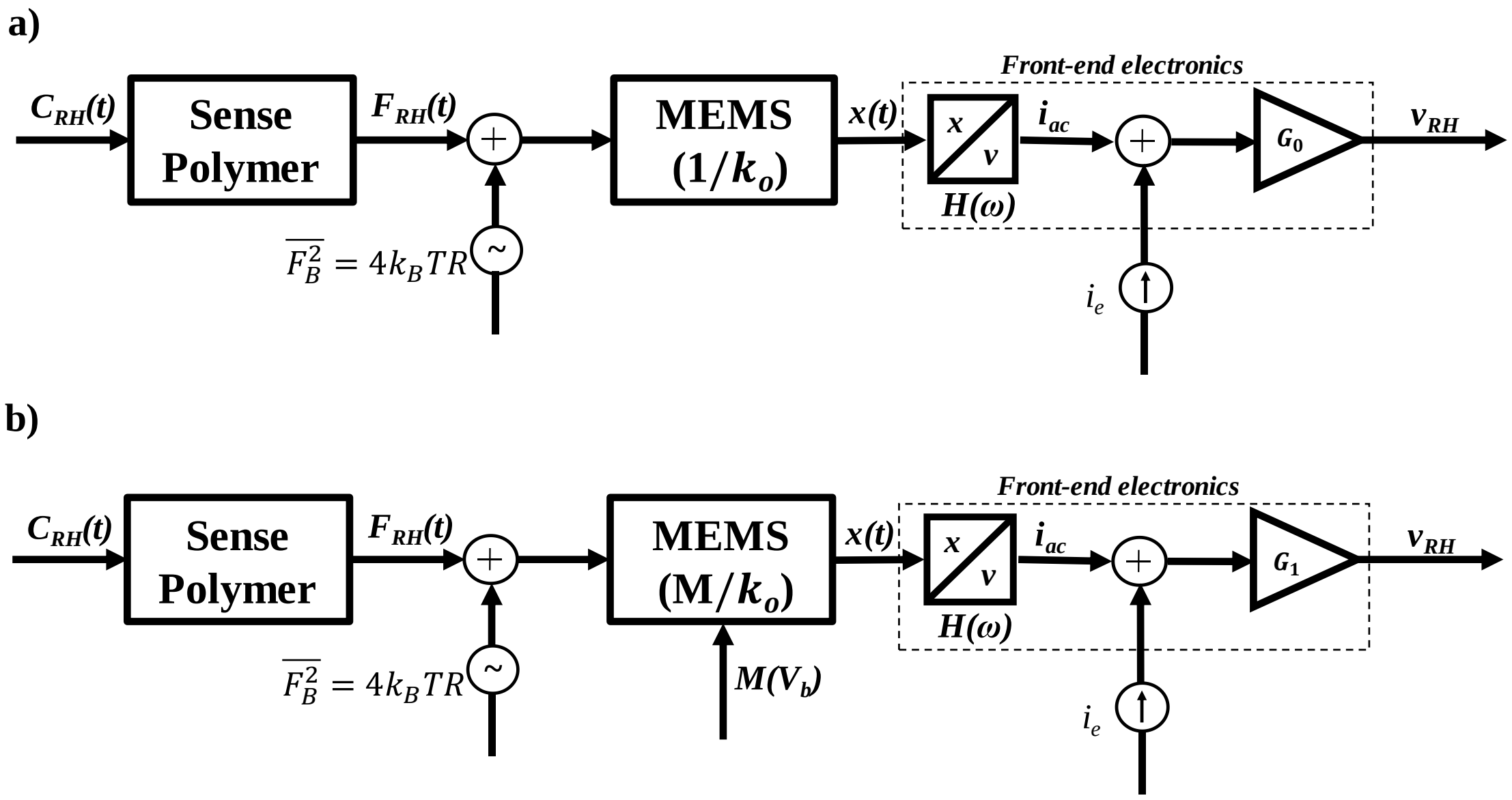

**Figure 3:** a) Schematic of the sensor system at zero bias with default spring constant $k_o$. b) With parametric amplification (near-Brownian noise limited)

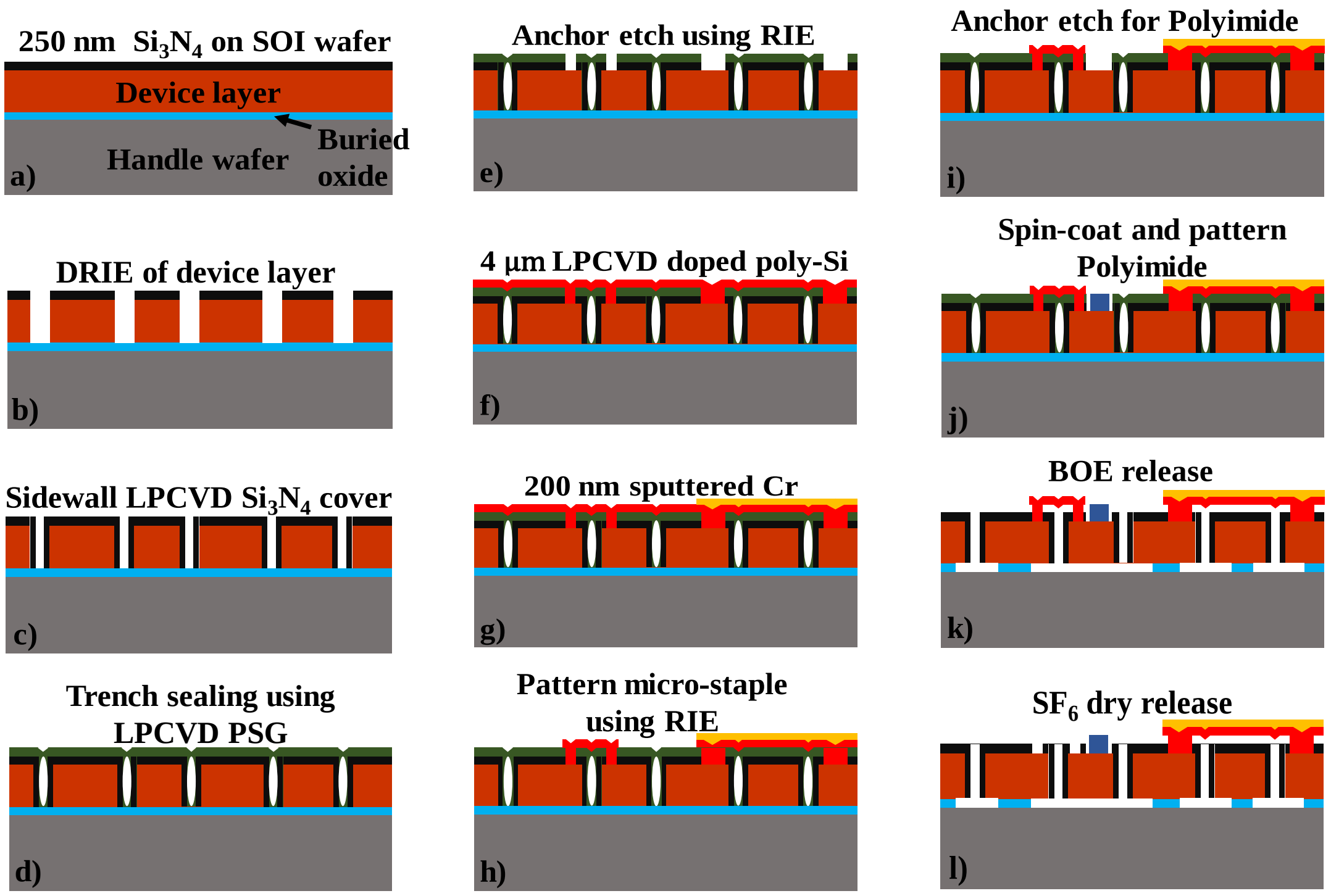

**Figure 4:** Simplified fabrication procedure of the device.

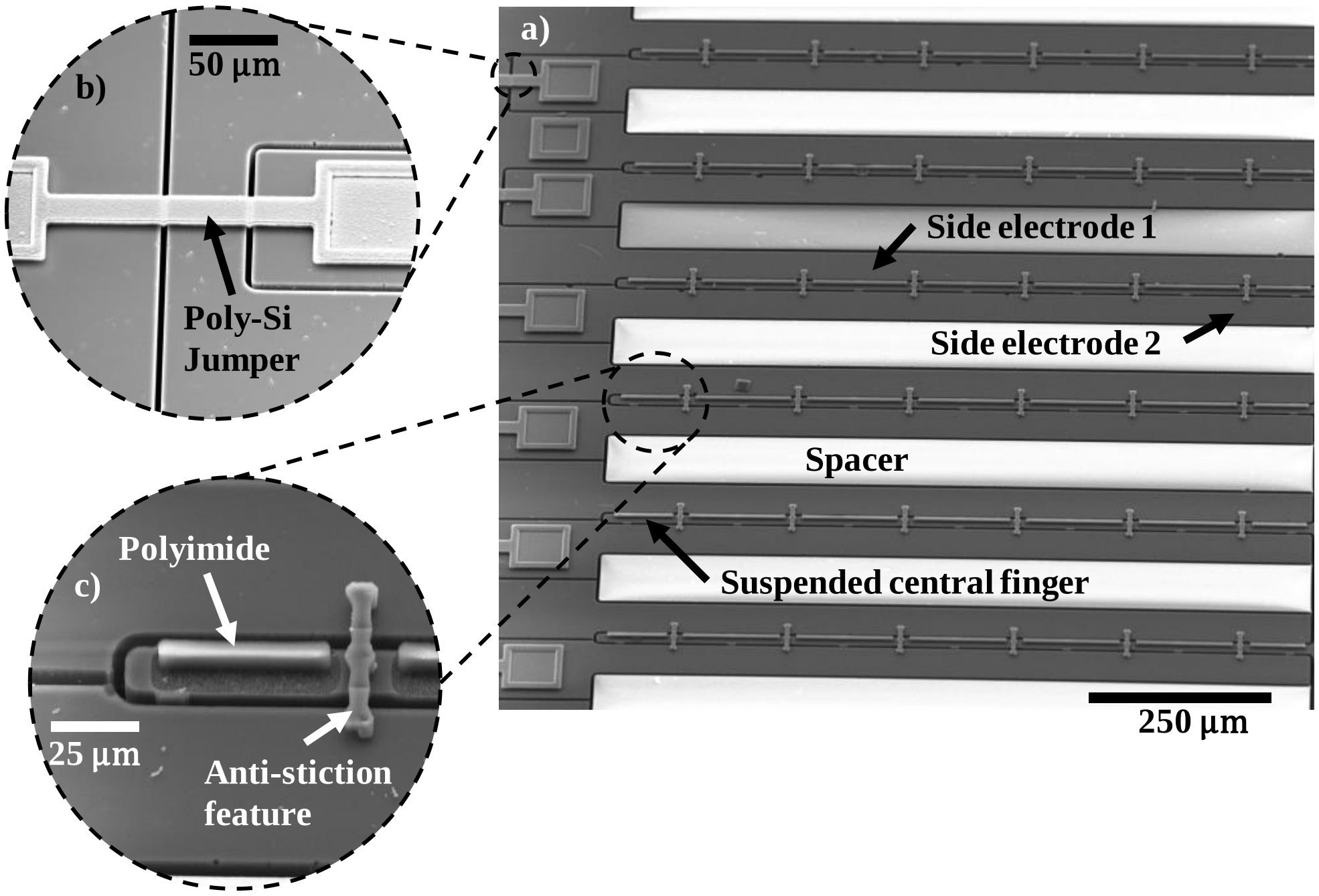

**Figure 5:** a) LACM sensor array b) Zoomed-in image of a poly-Si jumper over etched trenches acting as the 2nd level of electrical connections c) Magnified view of a suspended central finger with asymmetrically patterned polyimide and anti-stiction micro-staple holder.

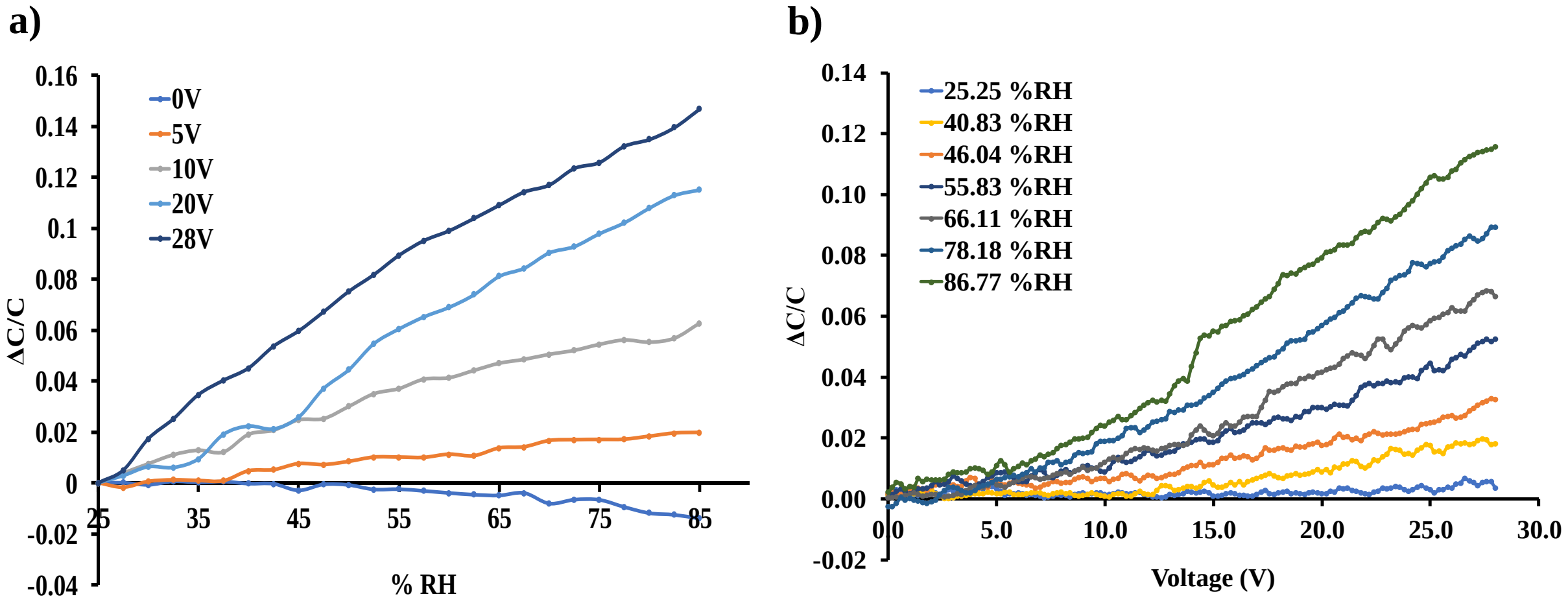

**Figure 6:** a) Sensor response to varying %RH levels at different DC bias voltage b) Sensor response to a varying bias voltage at different %RH levels.

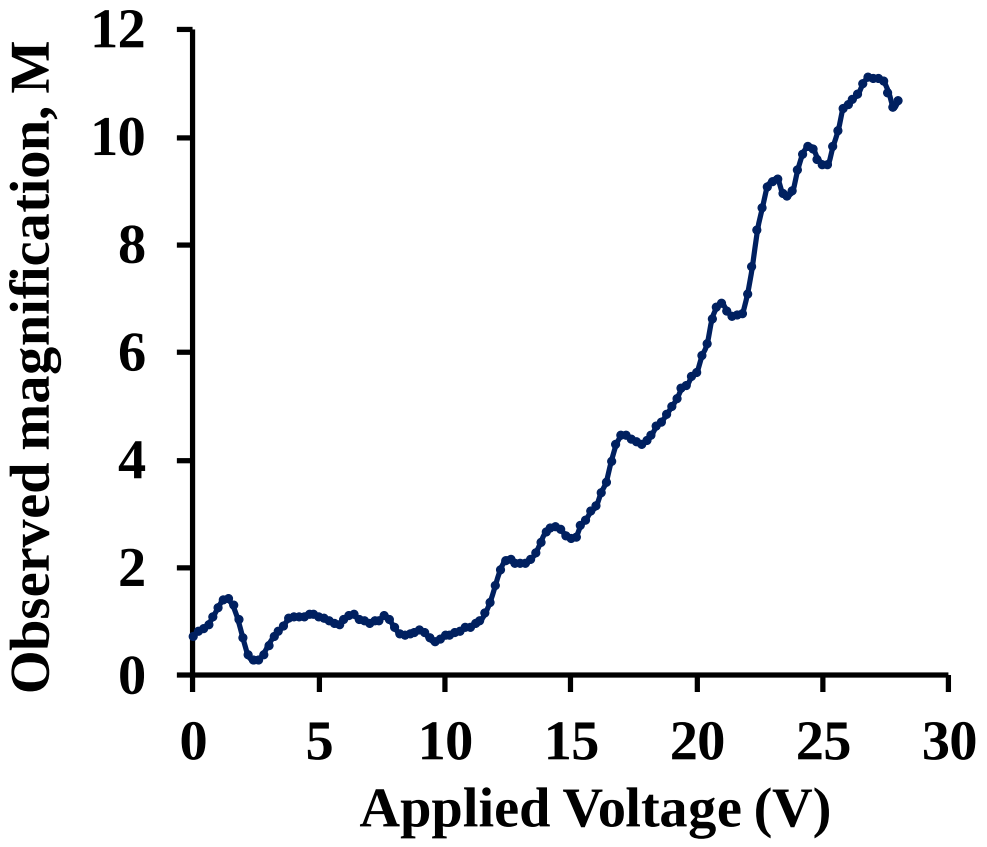

**Figure 7:** Observed sensor output magnification as a function of applied bias voltage at constant relative humidity.

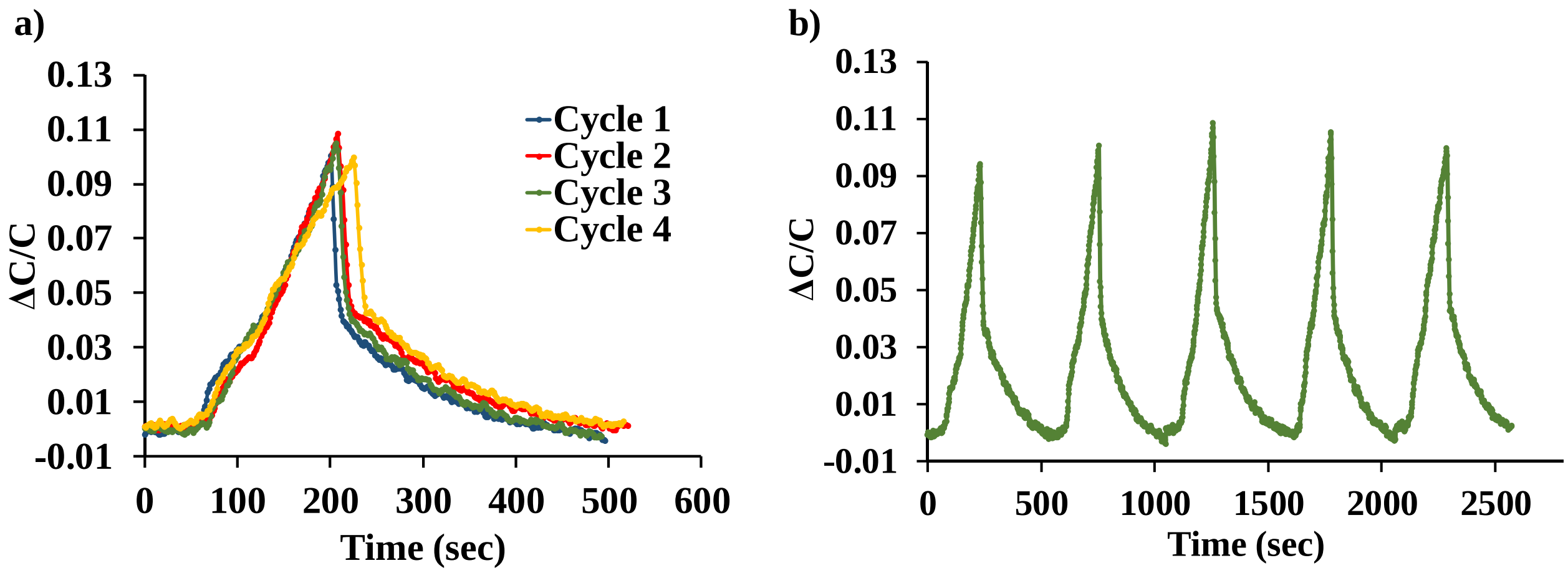

**Figure 8:** a) Repeatability of LACM sensor tested over four cycles. b) Sensor output over five consecutive cycles of exposure and removal of humidity.

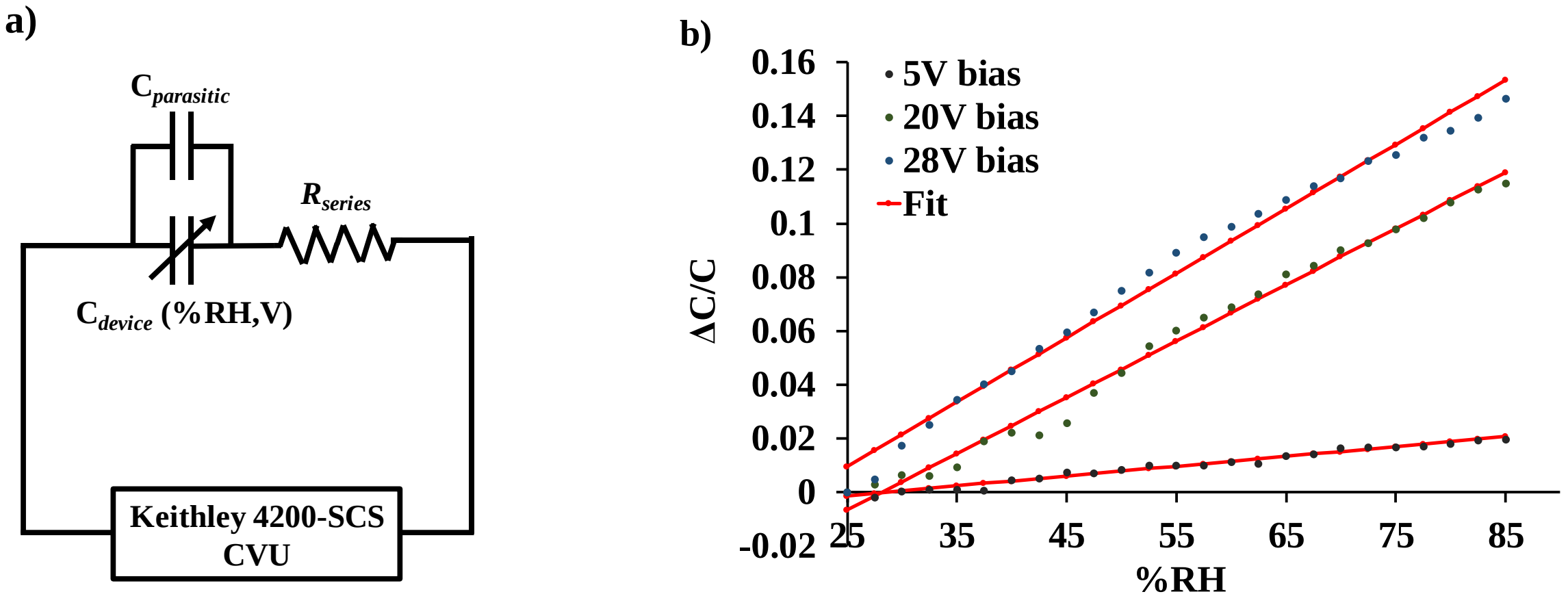

**Figure 9:** a) Equivalent electrical representation of the LACM sensor b) Normalized change in capacitance of the sensor curve fitted to equation (13).

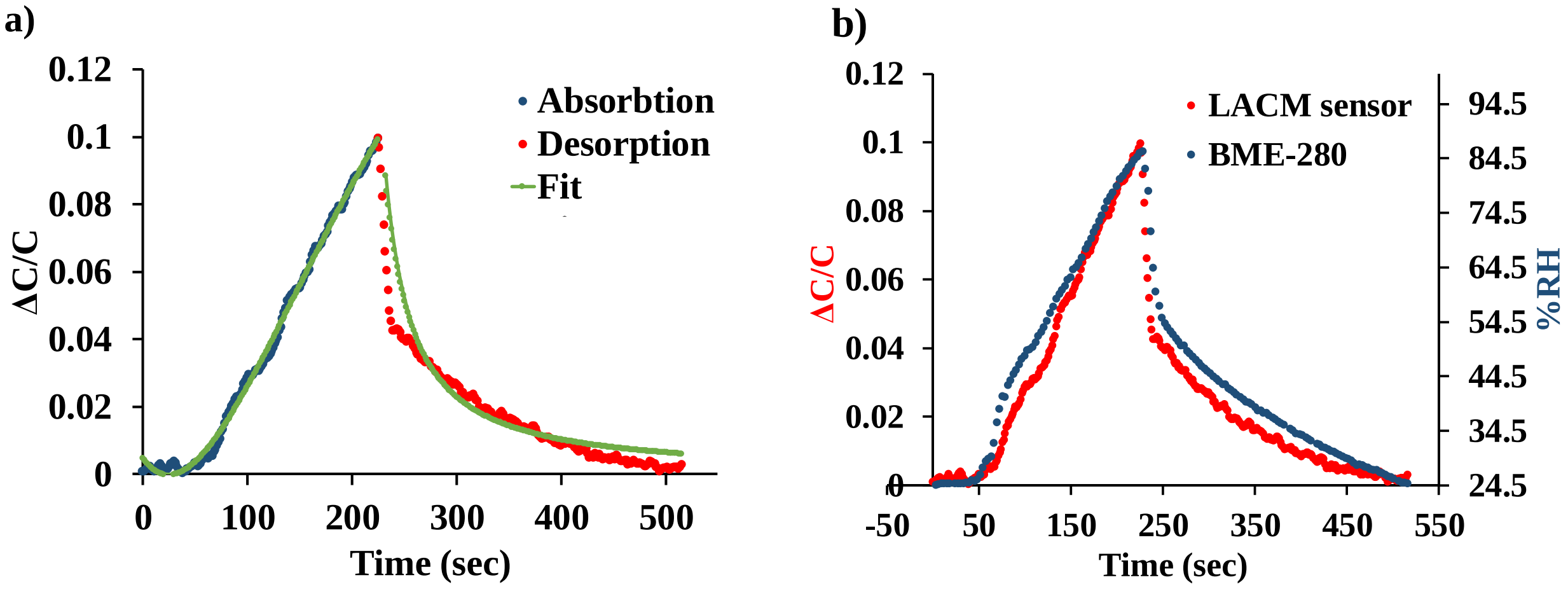

**Figure 10:** a) Dynamic response of the LACM sensor curve fitted to Fick's second law and the Polanyi-Wigner equation. b) Comparison of the LACM sensor performance to a commercially available BME-280 reference chip